\documentclass[letterpaper]{JHEP3}

\pdfoutput=1

\newcommand{\tr}{\mathrm{Tr}}

\usepackage{amsfonts}
\usepackage{amsmath}
\usepackage{cite}

\title{On black hole thermodynamics from super Yang-Mills}

\author{Toby Wiseman\\
{\it Theoretical Physics Group, Blackett Laboratory, Imperial College, London SW7 2AZ, UK } \\
\email{t.wiseman@imperial.ac.uk}
}

\preprint{arXiv:1304.3938}

\date{April 2013}

\abstract{

We consider maximally supersymmetric $U(N)$ Yang-Mills in $(1+p)$-dimensions for $p < 3$. In the 't Hooft large $N$ limit this is conjectured to be dual to $N$ Dp-branes in the decoupling limit. At low temperatures $T \ll \lambda^{1/(3-p)}$ governed by the dimensionful 't Hooft coupling $\lambda$, supergravity black holes predict the free energy density goes as $\sim N^2 T^{\frac{2(7-p)}{(5-p)}}$ and the expectation value of the scalars goes as $\sim T^{\frac{2}{5-p}}$, with dimensions made up by $\lambda$. 
The purpose of this work is to explain the origin of these peculiar powers of temperature.
We argue that these powers naturally arise by requiring that the low energy moduli of the theory become strongly coupled at low temperature.
As an application, we consider the BMN quantum mechanics that results from a supersymmetric deformation of the $p=0$ theory. The black holes dual to this deformed theory have not yet been constructed, and our analysis can be used to make an explicit prediction for their thermodynamic behaviour.

}

\begin{document}

%
\section{Introduction}
%

Maldacena's AdS-CFT duality \cite{Maldacena:1997re} has remarkable potential to allow us to understand the most basic questions in quantum black hole physics. 
The simplest realisations from the field theory side of the correspondence are those arising from maximally supersymmetric Yang-Mills (MSYM) in $(1+p)$-dimensions which is dual to the decoupling limit of Dp-branes \cite{Itzhaki:1998dd}, the original AdS-CFT case being $p=3$. 
Whilst these field theories are straightforward to write down and are well defined for $p \le 3$, it has so far not been possible to solve them. In the cases $p <3$ the gauge theory is strongly coupled in the IR and for  the conformal $p=3$ case, the dimensionless coupling must be taken to be large in order to achieve a weakly curved gravity dual. The challenge is to directly solve the thermal physics of these strongly coupled gauge theories and extract the dual quantum behaviour of black holes. 
\footnote{
For D1-D5 brane systems one obtains a realisation of AdS-CFT where the CFT is $(1+1)$-dimensional \cite{Aharony:1999ti}. In these cases the entropy of the dual BTZ black holes can be exactly reproduced using CFT methods\cite{Strominger:1996sh}. It is worth emphasising that whilst this is a great success, the CFT is not directly solved. It is only the entropy that can be computed, and other quantities cannot be solved for. 
}

A fascinating mean field approach was developed by Kabat, Lifschytz and Lowe \cite{Kabat:1999hp,Kabat:2000zv,Kabat:2001ve} for the $p=0$ case to directly solve the thermal theory, and extract the thermodynamics, with some success. Very recently there has been exciting renewed efforts to develop this method further \cite{Lin:2013jra}. 
Beyond these works there has been a distinct lack of analytic progress on explicit calculation of the thermal theory for any $p\le 3$. 
Over the past decade an exciting new direction has been the direct numerical Monte Carlo simulation of these gauge theories at finite temperature on the lattice (or `non-lattice') in the large N limit relevant for Maldacena decoupling. While the conformal $p=3$ case is perhaps the most elegant to consider from an analytic perspective, from the gravitational point of view the quantum black hole physics is essentially the same in all the cases of $p \le 3$. For the numerical approach lower dimensions is a big advantage, meaning less degrees of freedom to simulate and a simpler continuum limit. Indeed for $p=0$, the BFSS quantum mechanics \cite{Banks:1996vh}, the gauge theory is  a  quantum mechanics, hugely simplifying the continuum limit. In a lattice approach this implies that complicated supersymmetric lattice actions are not required \cite{Catterall:2007fp} and an elegant and powerful non-lattice or spectral approach \cite{Hanada:2007ti} can be taken since the gauge can be entirely fixed up to the Polyakov loop freedom. The first simulations of this $p=0$ have directly seen thermodynamic behaviour consistent with the dual predictions from supergravity \cite{Anagnostopoulos:2007fw,Catterall:2008yz,Hanada:2008gy,Catterall:2009xn}, 
and have even allowed the $\alpha'$ corrections to supergravity to be extracted \cite{Hanada:2008ez}. The BMN \cite{Berenstein:2002jq} deformation of the $p=0$ quantum mechanics, which we discuss later, has been studied \cite{Catterall:2010gf}. Going beyond $p=0$, the first thermal simulations of the $p=1$ case on a spatial circle \cite{Catterall:2010fx} have seen evidence of an expected Gregory-Laflamme phase transition \cite{Gregory:1993vy,Susskind:1997dr,Barbon:1998cr,Li:1998jy,Aharony:2004ig}. 
The higher $p$ cases are likely to be practical in the future, with the first work appearing recently on the conformal $p=3$ case \cite{Catterall:2012yq}. 
For excellent recent reviews on this direct Monte Carlo approach see \cite{Nishimura:2012xs,Hanada:2012eg}.\footnote{A number of other numerical works have simulated supersymmetric gauge theories to test holography and Matrix theory in the zero temperature context \cite{Hiller:2000nf,Campostrini:2004bs,Hiller:2005vf,Hanada:2009ne,Hanada:2011fq}.}
Beyond equilibrium thermodynamics an exciting numerical approach has been employed \cite{Asplund:2011qj,Riggins:2012qt,Asplund:2012tg} to study dynamics of initial states that thermalize and to understand how information is `scrambled' \cite{Sekino:2008he}. 

In the case of $p=3$ the conformal invariance specifies that energy density $\epsilon \sim N^2 T^4$ for temperature $T$. 
For the cases of $p < 3$ there is more texture in the thermodynamics.\footnote{
In fact for $p<3$ the canonical ensemble does not exist in these cases as the Dp-branes may Hawking evaporate from the decoupling region. This Hawking radiation rate is subdominant in $N$, and cannot be seen directly in the supergravity Dp-brane black hole thermodynamics, and at large $N$ we may think of there being a metastable thermal state. More formally, addition of a finite but small mass term cures this IR divergence of the canonical partition function, and in the limit of small mass will not change the leading $N$ behaviour.
 For an extended discussion of this in the context of $p=0$ see \cite{Catterall:2009xn} where the consequences for Monte-Carlo simulation are discussed. For a recent estimate of this Hawking radiation rate in the string dual see \cite{Lin:2013jra}. 
}
Lacking conformal symmetry, the 't Hooft coupling $\lambda = N g_{YM}^2$ is dimensionful and one may form the dimensionless temperature $t \equiv T / \lambda^{1/(3-p)}$. Now the energy density may have an interesting dependence on this dimensionless parameter. For $p < 3$ and high temperatures, $t \gg 1$, these theories are weakly coupled, and hence have energy density going as,
$
\epsilon \sim N^2 T^{1+p} 
$,
being simply determined by the number of UV degrees of freedom in the theory.
However, in the opposite low temperature limit, where as $N \to \infty$ then $t$ is finite but small, so $t \ll 1$,
the theory is strongly coupled and a dual supergravity description in terms of a Dp-brane black hole becomes valid, predicting a particular N dependence and power law scaling with temperature as, 
\begin{eqnarray}
\epsilon \sim N^2 t^{\frac{2 (7-p)}{(5-p)}} \lambda^{\frac{1+p}{3-p}} \; , \qquad t \ll 1 \; .
\end{eqnarray}
It is then a challenge for direct tests, numerical or otherwise, to reproduce both the correct temperature power law dependence, and also the overall coefficient. As mentioned above in the case of $p=0$ this has been achieved with great success using the direct numerical Monte Carlo approach.

One might hope that just as for the conformal $p=3$ case, the power law dependence of energy density with temperature can be analytically understood directly from gauge theory, even if calculating the precise coefficient cannot be achieved. It is the purpose of this paper to confirm this expectation. 
Our principle motivation is that so far the direct approaches (mean field or Monte Carlo) to solving $p<3$ have not \emph{assumed} this power law dependence, but rather have tried to \emph{reproduce} it. However, clearly if it has a simple analytic origin in the gauge theory, one might hope to significantly improve these direct approaches by building in this analytic knowledge. For example, in the numerical approach one might hope this would yield understanding of the most relevant field configurations and could be used to improve the Monte Carlo sampling. For the mean field approach one might develop better informed Gaussian approximations to expand about. 

Our starting point is the work of Smilga, who in the case of $p=0$ considers the dynamics of the low energy moduli of the theory \cite{Smilga:2008bt}. Smilga considers the quantum effective action of these moduli. The 1-loop correction to the moduli action is well known (the famous $v^4/r^7$ term in Matrix theory \cite{Banks:1996vh,Okawa:1998pz}) and Smilga supposes that the configurations where the tree level and 1-loop terms are of the same order, ie. the moduli are strongly coupled, reproduce the correct temperature scaling. Making certain assumptions, the strongest being to assume the correct $N^2$ dependence, so taking $\epsilon \sim N^2 t^q \lambda^{1/3}$ for some $q$, Smilga then deduces the correct power $q = 14/5$ that arises from the dual gravity by equating the leading tree level and 1-loop moduli action terms. Hence the peculiar dynamical scaling of energy density with temperature is given an origin within the MSYM. 
This approach is highly reminiscent of the work of Horowitz and Martinec \cite{Horowitz:1997fr}, Li  \cite{Li:1997iz}, and Banks, Fischler, Klebanov and Susskind  \cite{Banks:1997tn} in the Matrix theory context to explain the parametric dependence of energy with horizon radius for Schwarzschild black holes using precisely the same logic, namely taking the moduli action to be strongly coupled by equating the tree and 1-loop terms in the moduli action. 
%
%
Equating tree and 1-loop terms has also been recently used in \cite{Iizuka:2013yla} to explore the fascinating dynamics of the dual black hole formation in these MSYM theories.

Motivated by these earlier Matrix theory works, and Smilga's application of these ideas to the Maldacena limit of D0-branes, we may ask various questions;
\begin{itemize}
\item In the 't Hooft limit with finite but small $t \ll 1$ can we explain the temperature power law dependence $\epsilon \sim N^2  t^{\frac{2 (7-p)}{(5-p)}} \lambda^{\frac{1+p}{3-p}}$ for all $p < 3$, not just $p=0$? 
\item Can we explain it without assuming the correct $N$ dependence (as Smilga assumed in \cite{Smilga:2008bt})? 
\item In addition to the energy density, Maldacena duality \cite{Itzhaki:1998dd} predicts the behaviour of the scalar vevs to have a particular power law dependence on temperature. Can we predict this too?
\item The power law dependence $\epsilon \sim N^2 t^{\frac{2 (7-p)}{(5-p)}} \lambda^{\frac{1+p}{3-p}}$ only arises when the dual supergravity is valid, at low temperatures $t \ll 1$. Indeed, as we stated above, at high temperature $\epsilon \sim N^2 T^{1+p}$ for the cases $p < 3$. Can we understand why this temperature scaling is only obtained in this particular low temperature limit (by which we mean $t$ finite but small as $N \to \infty$)? 
\end{itemize}

In this paper we shall address these questions. We will consider the thermal effective action for the moduli of the $p < 3$ theories. We will show that requiring the moduli action becomes strongly coupled, following the ideas of \cite{Horowitz:1997fr,Li:1997iz,Banks:1997tn,Smilga:2008bt}, naturally yields
both the correct $N$ and $t$ dependence predicted by the dual Dp-brane black holes for both the energy density and the scalar expectation values for all $p < 3$. In addition we will explain why these $t$ scalings only arise at low temperature where $t$ is finite but $t \ll 1$. 

Beyond explaining the origin of the power law temperature scaling of the energy density and scalar vevs predicted by the usual dual Dp-brane black holes, these ideas may be used to make quantitative predictions for supergravity black hole solutions that have not yet been found. In particular in the case of $p=0$ MSYM one may deform this BFSS gauged quantum mechanics preserving the 16 supercharges and introduce a mass parameter $\mu$, to give the BMN quantum mechanics \cite{Berenstein:2002jq}. 
As emphasised in \cite{Catterall:2009xn} this is a particularly attractive system to study directly by Monte Carlo methods, as it is low dimensional, it has a IIA supergravity dual even with the mass deformation, and the addition of the mass naturally cures the IR singularities of the $p<3$ theories hence rendering the canonical partition function finite. Despite this fact, there has been relatively little work on it, the first Monte Carlo study being \cite{Catterall:2010gf}.
The zero temperature IIA supergravity duals are known from the work of Lin \cite{Lin:2004kw} and Lin, Lunin and Maldacena \cite{Lin:2004nb}, but finite temperature black holes with the correct UV asymptotics are not yet known, and this remains an important outstanding gravitational problem that must be solved to provide predictions for direct tests on the MSYM side.
We argue that in the regime where supergravity is good, our simple analysis strongly constrains the dependence of the energy density and scalar vevs on the temperature and mass $\mu$. Since we do not yet know the gravity dual solutions, these constraints may be regarded as a gauge theory \emph{prediction} for supergravity.

The structure of this paper is as follows.
In the next section we will briefly review the supergravity predictions for the energy density and scalar behaviour at low temperature for $(1+p)$-dimensional MSYM. Then in the following section \ref{sec:argument} we will give the simple argument to explain the origin of the peculiar power law dependence on temperature in these quantities. We will consider the quantum corrections to the low energy moduli space of these theories, and requiring these to be strongly coupled at low temperature will reproduce the correct power laws. For readability, we will not discuss the somewhat technical derivation of these quantum corrections until the later section \ref{sec:calc} of the paper.
In section \ref{sec:BMN} we will discuss the BMN deformation of the $p=0$ theory, and using the same arguments, deduce constraints on the low temperature behaviour of the thermodynamics and scalar vevs for the, as yet unknown, supergravity black holes. In section \ref{sec:calc} we will give the deferred details of the calculations of the thermal quantum 1-loop corrections to the classical moduli action. We conclude with a brief discussion.

%
\section{MSYM and supergravity black holes}
\label{sec:sugra}
%

We begin with maximally supersymmetric $U(N)$ Yang-Mills theory with coupling $g_{YM}$. In the large $N$ 't Hooft limit the natural coupling becomes $\lambda = N g_{YM}^2$. The bosonic part of the action is given as,
\begin{eqnarray}
\mathcal{S}_{B} = \frac{N}{\lambda}  \int dt dx^p \, \tr\left[  - \frac{1}{4} F_{\mu\nu}^2 - \frac{1}{2} D^\mu \Phi^I D_\mu \Phi^I + \frac{1}{4} \left[ \Phi^I , \Phi^J \right]^2  \right]
\end{eqnarray}
and the fermionic action is,
\begin{eqnarray}
\mathcal{S}_{F} = -\frac{1}{2} \frac{N}{\lambda} \int dt dx^p\, \tr \, \bar{\Psi} \left( \gamma^\mu D_\mu - i  \left[ \gamma^I \Phi^I , \cdot \right] \right) \Psi
\end{eqnarray}
where $x^\mu$ with $\mu = 0, \ldots, p$ are the world volume coordinates with time $x^0 = t$ and spatial coordinates $x^i$ with $i=1,\ldots,p$. Then $\Phi^I$ with $I = 1+p, \ldots, 9$ are the spacetime scalars representing the transverse degrees of freedom of the branes. They are $N \times N$ Hermitian matrices transforming in the adjoint of the gauge group. The fermion $\Psi$ is a $(1+9)$-dimensional Majorana-Weyl spinor and also transforms in the adjoint, and the collection $\{ \gamma^\mu, \gamma^I \}$ are the appropriate $SO(1,9)$ gamma matrices. As usual, $F_{\mu\nu} = \partial_\mu A_\nu - \partial_\nu A_\mu + i [ A_\mu , A_\nu ]$, and $D_\mu = \partial_\mu - i [ A_\mu, \cdot]$, with $A_\mu$ the gauge field potential transforming in the adjoint. 

Maldacena duality \cite{Itzhaki:1998dd} states that at finite temperature there is a dual closed IIA ($p$ even) or IIB ($p$ odd) string theory description of this gauge theory in terms of the decoupling limit of thermal Dp-branes. In the large $N$ 't Hooft limit this string theory description may reduce to one in IIA or IIB supergravity, where the string frame metric takes the form,
\begin{eqnarray}
ds^2 & = & \alpha' \left( \frac{U^{\frac{7-p}{2}}}{\sqrt{a_p \lambda}} (-f dt^2 + dx^{i} dx^i) + \sqrt{a_p \lambda} \left( U^{-\frac{7-p}{2}} \frac{dU^2}{f} + U^{\frac{p-3}{2}} d\Omega^2_{(8-p)}\right) \right) \nonumber \\
f(U) & = & 1 - \left( \frac{U_0}{U} \right)^{7-p} \; , \qquad a_p = 2^{7- 2 p} \pi^{\frac{9-3 p}{2}} \Gamma\left( \tfrac{7-p}{2} \right)
\end{eqnarray}
and the dilation goes as,
\begin{eqnarray}
\label{eq:dilaton}
e^\phi =  \frac{1}{N}  ( 2 \pi )^{2-p} ( a_p )^{\frac{3-p}{4}} \left( \frac{U}{\lambda^{1/(3-p)}}   \right)^{- \frac{(3-p)(7-p)}{4}}  
\end{eqnarray}
and in addition the $(1+p)$-form potential carries the charge of the $N$ Dp-branes. Here $t$ and $x^i$ are the Dp-brane $(1+p)$-dimensional world volume coordinates. The radial coordinate $U$ is identified with an energy scale associated to the expectation value of the scalars in the MSYM. The horizon is at $U = U_0$, implying the thermal scale for the scalars is $U_0$
 so that,
\begin{eqnarray}
U_0 \sim \sqrt{  \frac{1}{N} \langle \tr \Phi^I \Phi^I \rangle }
\end{eqnarray}
at finite temperature.

In order for the supergravity solution to consistently describe the full closed string theory physics we require that both stringy and $\alpha'$ corrections are small. Stringy corrections are characterised by the string coupling, the dilaton. The $\alpha'$ corrections are characterised by the size of the curvature of the solution relative to the string scale set by $\alpha'$. The radius of curvature $R$ goes as,
$
\alpha' /R^2 \sim \left( U / \lambda^{1/(3-p) } \right)^{\frac{3-p}{2} } 
$. 
We will consider the large $N$ 't Hooft limit, where it is natural to take,
\begin{eqnarray}
N \to \infty \; , \qquad \frac{U}{\lambda^{1/(3-p)}}  \; , \; \frac{U_0}{\lambda^{1/(3-p)}}  \sim \mathrm{finite}
\end{eqnarray}
and we see from \eqref{eq:dilaton} that in this large $N$ limit the solution is well described by semiclassical string theory since the dilaton will everywhere be small. However, in order to ensure that supergravity gives a good description we shall in addition require that the curvature $\alpha'$ corrections are small, and so we must take,
\begin{eqnarray}
U  \; , \; U_0 \ll \lambda^{1/(3-p)} \; .
\end{eqnarray}
We emphasise that we require these quantities to be small, but \emph{finite}, in the large $N \to \infty$ limit. 

The expectation value for the energy density $\epsilon$ of MSYM is given by the energy density of the Dp-brane solution above extremality, and yields, \begin{eqnarray}
\frac{  \epsilon  }{ \lambda^{(1+p)/(3-p)} } = b_p N^2  \left( \frac{ U_0 }{ \lambda^{1/(3-p)} } \right)^{7-p} \; , \quad 
b_p = \frac{8 (7-p) \pi^2}{9-p} a_p 
\end{eqnarray}
as deduced either from the Dp-brane solutions before taking the decoupling limit, or else from holographic renormalization that has been developed for these non-conformal Dp-brane geometries \cite{Wiseman:2008qa,Kanitscheider:2008kd}. Defining the dimensionless temperature $t = T / \lambda^{1/(3-p)}$ then semiclassical black hole thermodynamics yields,
\begin{eqnarray}
\frac{ U_0 }{ \lambda^{1/(3-p)} }  =  c_p \, t^{\frac{2}{5-p}} \; , \quad c_p =  \left( \frac{16 \pi^2 a_p}{(7-p)^2} \right)^{\frac{1}{5-p}}
\; .
\end{eqnarray}
The condition $U_0  \ll \lambda^{1/(3-p)}$ above for supergravity to be valid translates into the condition,
$
t \ll 1
$
on the dimensionless temperature. Thus we see that gravity makes the following parametric predictions at low temperature;
\begin{eqnarray}
 \epsilon \sim N^2 t^{\frac{2(7-p)}{5-p}}  \lambda^{(1+p)/(3-p)} \; , \quad \sqrt{  \frac{1}{N} \langle \tr \Phi^I \Phi^I \rangle } \sim U_0 \sim t^{\frac{2}{5-p}} \lambda^{1/(3-p)}
\end{eqnarray}
for $t \ll 1$.

%
\section{The basic argument}
\label{sec:argument}
%

We wish to consider putting our MSYM at finite temperature $T = 1/\beta$. It is convenient then to use the Euclidean time formulation, where Euclidean time $\tau = i \, t$ is periodic with period $\beta$, and the action $S = i S^E$ with $S^E$ the Euclidean action, and $A^\tau = i A^t$ for the gauge field. Hence now $x^\mu = ( \tau, x^i )$. It is also convenient to introduce vector notation for the $SO(9-p)$ symmetry index, so now $\Phi^I \to \vec{\Phi}$.

A set of classical vacua of this theory are gauge equivalent to the configurations,
\begin{eqnarray}
A^\mu_{ab} =  a^\mu_a \delta_{ab}  \; , \quad \vec{\Phi}_{ab} =  \vec{\phi}_a \delta_{ab} 
\end{eqnarray}
for real constants $a^\mu_a$ and $\vec{\phi}_a$,
with fermion fields being trivial, and $a,b = 1, \ldots , N$ are the colour indices of the adjoint hermitian matrices. 
Such a configuration
breaks the $U(N)$ gauge symmetry to $U(1)^{N}$. The constants $a^\mu_a$ then represent constant gauge field configurations for this $U(1)^N$ gauge symmetry.
The $\vec{\phi}_a$ give the transverse displacement of the $N$ Dp-branes. 
We may promote the constants $a^\mu_a$, $\vec{\phi}_a$ to slowly varying real scalar moduli fields, $a^\mu_a(\tau,x)$, $\vec{\phi}_a(\tau, x)$, and then these have a classical moduli space Euclidean action,
\begin{eqnarray}
\label{eq:moduli}
{S}^{E,classical} =  \frac{N}{\lambda} \int d\tau dx^p  \sum_{a}  \left( \frac{1}{2} \partial^\mu \vec{\phi}_a\cdot\partial_\mu \vec{\phi}_a + \frac{1}{4} F_{\mu\nu a} F^{\mu\nu}_a \right) \; .
\end{eqnarray}
where $F_{\mu\nu a} = \partial_\mu a_{\nu a} - \partial_\nu a_{\mu a}$ are the field strengths for the $N$ $U(1)$ gauge fields.
Considering non-zero configurations for these moduli fields implies that the off diagonal components of the bosonic matrices $A^\mu$, $\vec{\Phi}$ and fermionic spinor matrices $\Psi$ are massive degrees of freedom. The $ab$ components of these matrices have mass squared $\sim | \vec{\phi}_a - \vec{\phi}_b |^2$. Considering configurations of the moduli such that all these masses are large, one may integrate out the off-diagonal degrees of freedom to yield quantum corrections to this classical moduli space action. Whilst with maximal supersymmetry there can be no corrections to the potential and kinetic terms, in fact one does find higher derivative quantum corrections. 

We now apply a simple extension of the standard calculations familiar from consideration of Matrix theory \cite{Banks:1996vh,Okawa:1998pz,Ambjorn:1998zt} to deduce various 1-loop quantum corrections to this leading tree level moduli space action. We will study these quantum corrections up to 4 derivatives in the scalars, and ignore terms involving derivatives of the gauge field moduli. At 1-loop the terms come in two varieties; those that are independent of temperature (and hence are generated at zero temperature), and those that explicitly depend on temperature. We denote these the \emph{non-thermal} and \emph{thermal} 1-loop corrections respectively. We will now state the relevant results here, and defer their calculation until the later section \ref{sec:calc}.

Firstly we consider the non-thermal correction.
Up to 4 derivatives in the scalars, the non-thermal 1-loop correction has a single contribution of the form,
\begin{eqnarray}
\label{eq:quantcorr}
{S}^{E,1-loop}_{non-thermal} &=&  - \int d\tau dx^p \sum_{a<b}  \frac{\Gamma\left( \frac{7 - p}{2} \right)}{( 4 \pi )^\frac{1+p}{2}} \Bigg( 
2 \frac{ \left( \partial_\mu \left( \vec{\phi}_a - \vec{\phi}_b \right) \cdot \partial_\nu \left( \vec{\phi}_a - \vec{\phi}_b \right) \right) \left( \partial^\mu \left( \vec{\phi}_a - \vec{\phi}_b \right) \cdot \partial^\nu \left( \vec{\phi}_a - \vec{\phi}_b \right) \right) }{ | \vec{\phi}_a - \vec{\phi}_b |^{7-p} } 
\nonumber \\
&& \qquad \qquad \qquad \qquad
\qquad \qquad \qquad \qquad
-  \frac{\left( \partial_\mu \left( \vec{\phi}_a - \vec{\phi}_b \right) \cdot \partial^\mu \left( \vec{\phi}_a - \vec{\phi}_b \right)  \right)^2 }{ |  \vec{\phi}_a - \vec{\phi}_b  |^{7-p} }
\Bigg)  + \ldots
\end{eqnarray}
where supersymmetry ensures that the potential and 2 derivative kinetic terms are uncorrected. 
We see that there is no explicit temperature dependence beyond the period of Euclidean time, and that the contribution, being negative, is attractive.
This term is the generalisation of the famous $v^4/r^7$ correction familiar from Matrix theory \cite{Banks:1996vh}. The dots $\ldots$ above are terms involving higher derivatives of the scalars, which on dimensional grounds take the schematic form, $\sim \left( (\partial \phi)^4 /  \phi^{7-p} \right) \left( (\partial \phi)^2 / \phi^4 \right)^{n}$, for integer $n \ge 1$ (see for example \cite{Becker:1997xw} for the case $p=0$). As we shall see later, these will be suppressed relative to the leading 4 derivative term.
In addition, recall that we are only keeping track of terms involving derivatives of the scalars, and we similarly expect 4-derivative $F^4/\phi^{7-p}$ corrections for the gauge fields \cite{Buchbinder:1999jn,Buchbinder:2001ui}.

In addition to these temperature independent terms, finite temperature breaks the supersymmetry and generates corrections for all even derivative orders that explicitly depend on temperature. However, an important feature of the supersymmetry is that due to the boson-fermion cancellations, these thermal corrections are exponentially suppressed in the limit,
\begin{eqnarray}
\label{eq:lowT}
\beta |  \vec{\phi}_{a} - \vec{\phi}_{b} | \gg 1 \; .
\end{eqnarray}
As we shall later see, it is precisely this limit that we will be interested in.
We find the leading 1-loop thermal correction in this limit to be,
\begin{eqnarray}
\label{eq:quantcorrthermal}
{S}^{E,1-loop}_{thermal} &=&  \int d\tau dx^p \sum_{a<b} \Bigg[  \frac{ U_a U^\star_b + U_b U^\star_a }{ \beta^{1 + p} }  e^{- \beta |  \vec{\phi}_{a} - \vec{\phi}_{b} |} \left( \beta |  \vec{\phi}_{a} - \vec{\phi}_{b} | \right)^{p/2}  \\
&& \qquad
\times \left( - \frac{16}{ (2 \pi)^{p/2} } + \mathrm{higher\;derivative\;terms} \right) \times \left( 1 + O\left( \frac{1}{ \beta |  \vec{\phi}_{a} - \vec{\phi}_{b} | } \right) \right)  \Bigg] \nonumber
\end{eqnarray}
where $U_a = e^{i \oint d a_a} = e^{i \oint d \tau a^0_a}$ are the Polyakov loops around the Euclidean time circle. In this limit 
$\beta |  \vec{\phi}_{a} - \vec{\phi}_{b} | \gg 1$ the two derivative corrections, being exponentially suppressed, are subdominant to the tree level kinetic two derivative terms in \eqref{eq:moduli}, and the four derivative corrections are subdominant to the non-thermal 1-loop four derivative term of \eqref{eq:quantcorr}. As we shall therefore not require details of these terms, for simplicity we have suppressed them in the above expression, giving only the zero derivative potential term which is the leading zero derivative correction to the moduli action.
This 1-loop zero derivative correction, the thermally generated potential, is the generalisation of that computed by Ambjorn, Makeenko and Semenoff in the case of $p=0$ \cite{Ambjorn:1998zt}.\footnote{
We also note that as discussed in \cite{Catterall:2009xn} for $p=0$, the 1-loop approximation becomes better in the regime $\beta |  \vec{\phi}_{a} - \vec{\phi}_{b} |\gg 1$ where for such a potential the moduli are essentially massless, and this leads to a divergence in the canonical partition function (at the subleading $O(N)$ rather than $O(N^2)$) as for $p<3$ one should perform an integration over these moduli fields (due to Coleman-Mermin-Wagner).} 
In the later section \ref{sec:calc} some of the higher derivative terms are calculated and their explicit form is given.

We now use an analogous approach to \cite{Horowitz:1997fr,Li:1997iz,Banks:1997tn,Smilga:2008bt} in order to deduce the $N$ and $T$ scaling of the quantities of interest. 
One important difference to these works is that we will consider the theory at finite temperature using the canonical Euclidean approach rather than a Hamiltonian Lorentzian point of view. This was first used in the Matrix theory context in \cite{Ohta:1998sq} and appears to have the virtue that the $N$ scaling need not be assumed, but emerges naturally.

Having given the 1-loop corrected moduli action, we now show that the peculiar parametric temperature dependence predicted by supergravity emerges precisely from requiring that the loop expansion becomes strongly coupled. Firstly we estimate all the $\vec{\phi}_a$ to be of the same order which we write as $\phi$. Furthermore we estimate that the differences between these moduli are also of the same scale, so that,
\begin{eqnarray}
\vec{\phi}_a \sim \vec{\phi}_a - \vec{\phi}_b \sim \phi \; .
\end{eqnarray}
%
%
At finite temperature Euclidean time is periodic, setting a length scale $\beta$. We now assume that it is this  scale that determines the behaviour of the derivatives, so that we estimate,
\begin{eqnarray}
\partial_\mu \vec{\phi}_a \sim  \partial_\mu \vec{\phi}_{a} - \partial_\mu \vec{\phi}_{b} \sim \frac{1}{\beta} \phi \; .
\end{eqnarray}
%
%
%
%
%
This is our key assumption - one might expect the quantum length scale $\lambda^{-1/(3-p)}$ and the dimensionless number $N$ to also play a role in determining derivative behaviour. While the estimate is geometrically natural, we do not have a strong justification for why the moduli derivative behaviour should only respond to the length scale $\beta$, but emphasise that the remainder of this argument hinges on this point. One possible justification is that in order to have rich dynamics at low temperatures the system must `forget' higher energy scales, and hence behaviour of derivatives will be determined (at least dimensionally) by the low energy scale $1/\beta$. This is in contrast to a theory with a mass gap (an example being the simple harmonic oscillator), whose thermodynamics at low temperature is trivial, with the dynamics collapsing onto the ground state, so that the energy scale $1/\beta$ does not influence the behaviour of the theory far below the gap scale. 

We now proceed to estimate when the loop expansion becomes strongly coupled, and do this by equating the classical moduli action for the scalars with the 1-loop correction, ignoring coefficients that are order one. We assume that the condition in equation \eqref{eq:lowT} holds, so that,
\begin{eqnarray}
\label{eq:assumption}
\beta |  \vec{\phi}_{a} - \vec{\phi}_{b} | \sim \beta {\phi} \gg 1
\end{eqnarray}
and then the 1-loop corrections are dominated by the non-thermal terms, with the thermal terms being exponentially suppressed. 
In addition it is worth emphasising that the 4 derivative 1-loop non-thermal term given in equation \eqref{eq:quantcorr} also dominates  all the higher derivative 1-loop non-thermal terms, since as discussed below that equation, these are related by increasing powers going as $( \partial \phi )^2 / \phi^4 \sim 1/(\beta \phi)^2$ and hence are power law suppressed relative to the 4 derivative term in this limit.
A key point is that we will shortly check the self-consistency of this assumption finding it holds precisely in the low temperature regime $t \ll 1$ dual to supergravity.

The classical moduli term is estimated as,
\begin{eqnarray}
\label{eq:classical}
 S^{E,classical} \sim \frac{N}{\lambda} \int d\tau dx^p \, \sum_{a} \left( \partial_\mu \vec{\phi}_a \right)^2 \sim   \int dx^p \, \frac{N^2}{\lambda \beta} \phi^2
\end{eqnarray}
where we use $\int d\tau \sim \beta$, $\sum_a \sim N$ and $\partial_\mu \vec{\phi}_a \sim \phi/\beta$. 
We similarly estimate the 1-loop contribution which is dominated by the 4 derivative non-thermal term to obtain,
 \begin{eqnarray}
 S^{E,1-loop} \sim  S^{E,1-loop}_{non-thermal} &\sim &\int d\tau dx^p \,  \sum_{a<b} \frac{ \left( \partial_\mu \left( \vec{\phi}_a -  \vec{\phi}_b \right) \right)^4 }{  | \vec{\phi}_{a} - \vec{\phi}_{b} |^{7-p} } \nonumber \\
& \sim & \int dx^p \,  \frac{N^2}{\beta^3 \phi^{3-p}}
 \end{eqnarray}
where we have taken, $\sum_{a<b} \sim N^2$. Now equating the classical and 1-loop terms yields the estimate,
\begin{eqnarray}
S^{E,classical} \sim S^{E,1-loop} \quad \implies \quad \phi   \sim  \left(  \beta \lambda^{\frac{1}{3-p}} \right)^{-\frac{2}{5-p}}  \lambda^{\frac{1}{3-p}} = t^{\frac{2}{5-p}}  \lambda^{\frac{1}{3-p}}
\end{eqnarray}
for the scalars. Having made the assumption \eqref{eq:assumption} which suppresses the 1-loop thermal corrections and higher derivative non-thermal corrections, we must now check it is consistent. Using our estimate,
\begin{eqnarray}
1 \ll  \beta \, \phi \sim \left( \beta \lambda^{\frac{1}{3-p}} \right)^{\frac{3-p}{5-p}} = t^{-\frac{3-p}{5-p}}
\end{eqnarray}
and hence we see for $p<3$ then our assumption is consistent when $t \ll 1$. Hence the above analysis is consistent precisely in the low temperature regime where supergravity black holes give the predictions we wish to derive. At higher temperatures one would have to consider the contribution from the 1-loop thermal terms and higher derivative non-thermal terms.

The thermal vev of the scalars is then estimated as,
\begin{eqnarray}
 \sqrt{ < \frac{1}{N} \tr \Phi^I \Phi^I > } & = &  \sqrt{ < \frac{1}{N} \sum_a | \vec{\phi}_a |^2 > } \sim \phi \sim  t^{\frac{2}{5-p}}  \lambda^{\frac{1}{3-p}} 
\end{eqnarray}
where again $\sum_a \sim N$,
and we recall that in the regime $t \ll 1$ where our analysis is self-consistent then supergravity predicts $\sqrt{ < \frac{1}{N} \tr \Phi^I \Phi^I > } \sim  t^{\frac{2}{5-p}} \lambda^{\frac{1}{3-p}}$ which exactly agrees with our estimate.
Encouraged by this we proceed to consider the free energy.
Now in this low temperature regime we may use \eqref{eq:classical} to estimate the classical and 1-loop terms which we have equated to give,
\begin{eqnarray}
S^E_{classical} , S^E_{1-loop}  \sim   \int dx^p \, \frac{N^2}{\lambda \beta} \phi^2 \sim \lambda^\frac{p}{3-p} \int dx^p N^2  \left( \beta \lambda^{\frac{1}{3-p}} \right)^{-\frac{9-p}{5-p}} \; .
\end{eqnarray}
%
%
The free energy, $F$, is estimated as $\beta F = - \ln Z \sim  S^E$. 
Hence we may make estimates of the free energy density $f$, and energy density $\epsilon$ (which for power law dependence on temperature has the same parametric dependence as $f$);
\begin{eqnarray}
\beta f \sim  \lambda^\frac{p}{3-p}  N^2  \left( \beta \lambda^{\frac{1}{3-p}} \right)^{-\frac{9-p}{5-p}} \quad \implies \quad
\epsilon \, , \; f \sim  N^2 t^{\frac{2(7-p)}{5-p}}  \lambda^\frac{1+p}{3-p} \; .
\end{eqnarray}
This is precisely the parametric dependence predicted by the supergravity in the regime $t \ll 1$ where our estimates are self-consistent.

Thus the condition that the quantum corrections to the classical moduli space action become strongly coupled at finite temperature precisely reproduces the supergravity predictions for the energy density and scalar vev in the regime $t \ll 1$. In this temperature range we see that the 1-loop quantum corrections are dominated by the 4-derivative \emph{non-thermal} terms (those with no explicit temperature dependence, that also are present at zero temperature), with the \emph{thermal} contributions (those with explicit temperature dependence) being exponentially suppressed relative to these, and the higher derivative non-thermal corrections being power law suppressed. At higher temperatures,  $t \sim 1$, the thermal 1-loop corrections and higher derivative non-thermal 1-loop terms are not suppressed relative to the leading non-thermal correction, and presumably the condition that the 1-loop corrections become strongly coupled is then much more complicated, and the simple power law temperature dependence of energy density and scalar vev will no longer apply. This is seen for $p=0$ Monte-Carlo simulations \cite{Anagnostopoulos:2007fw,Catterall:2008yz}.

In the next section we use the simple estimates above to make a prediction for the thermodynamic behaviour of certain black holes in IIA supergravity, those dual to the BMN quantum mechanics. Then the remainder of the paper, section \ref{sec:calc}, is spent computing the 1-loop quantum corrections given above in equations \eqref{eq:quantcorr} and \eqref{eq:quantcorrthermal}. 

%
\section{A prediction for black holes dual to the BMN quantum mechanics}
\label{sec:BMN}
%

Now we focus on the $p=0$ case, the BFSS quantum mechanics \cite{Banks:1996vh}, and add to the action a term $S_{BMN}$ that includes mass and Myers terms \cite{Myers:1999ps} that break the global $SO(9)$ symmetry to $SO(3) \times SO(6)$ whilst preserving the 16 supercharges of the theory \cite{Berenstein:2002jq}. This theory is the BMN quantum mechanics.  The bosonic part of this is,
\begin{eqnarray}
{S}^{bosonic}_{BMN} &=& \frac{N}{2 \lambda}  \int dt \, \tr \Bigg[ - \mu^2 \sum_{I=1}^{3} (\Phi^I)^2 - \frac{\mu^2}{4} \sum_{I=4}^{9} (\Phi^I)^2 -  2 i\, \mu \sum_{I,J,K=1}^{3}\epsilon_{IJK} \Phi^I \Phi^J \Phi^K  \Bigg] \; .
\end{eqnarray}
After this mass deformation of the BFSS theory, in the 't Hooft limit we have been considering there remains a IIA  string dual description.
The zero temperature IIA supergravity vacuum was given in \cite{Lin:2004kw,Lin:2004nb}. The finite temperature black holes in this asymptotic geometry have not yet been found. Since finite temperature breaks supersymmetry and the geometries are of cohomogeneity two this is likely to be solved only as a numerical problem.   
In order to ensure supergravity is valid we expect that one will require temperature $T \lambda^{1/3} \ll 1$ to avoid $\alpha'$ corrections near the horizon (as in the usual $p=0$ MSYM case), and in addition also  $\mu \lambda^{1/3} \ll 1$ for the same reason.
Thus there will be a one parameter family of solutions characterised by the dimensionless ratio $x \equiv \mu / T$.

In the supergravity limit for zero mass $\mu$ we know that the energy ($=$ energy density for $p=0$) and the scalar vev scale as,
\begin{eqnarray}
\epsilon = a \, N^2 t^{14/5} \lambda^{1/3} \; , \qquad  \sqrt{ \langle  \frac{ \tr \Phi^I \Phi^I }{N} \rangle }  = b \, t^{2/5}  \lambda^{1/3}
\end{eqnarray}
for $t \ll 1$ and constants $a$ and $b$. Adding the relevant BMN mass deformation, the question is now how these behaviours will be modified. Suppose we fix the dimensionless ratio $x = \mu / T$ and vary the temperature. We might expect energy and the scalar vevs again to have some power law scaling at low temperature. A priori one would expect that this power law would depend on the value of $x$, so that;
\begin{eqnarray}
 \epsilon &=& A(x) \, N^2 \, t^{p(x)} \lambda^{1/3} \; , \nonumber \\
  \sqrt{     \sum_{I=1,2,3} \langle\frac{ \tr \Phi^I \Phi^I }{N} \rangle } &=& B_1(x) \, t^{q(x)} \lambda^{1/3}  \; , \qquad \sqrt{  \sum_{I=4\, \ldots, 9}\langle \frac{ \tr \Phi^I \Phi^I }{N} \rangle } = B_2(x) \, t^{r(x)} \lambda^{1/3} 
\end{eqnarray}
again for $t = T / \lambda^{1/3}$, in the limit $T\, , \mu \ll \lambda^{1/3}$, so $x \sim O(1)$, where $A(0) = a$, $B_1(0) = B_2(0) = b$ and $p(0) = 14/5$, $q(0) = r(0) = 2/5$. We know of no argument from supergravity that would dictate the $x$ dependence of the leading powers $p(x), q(x), r(x)$ in this low temperature limit. 

Now consider the effect of the above mass deformation on our effective action for the moduli fields. The cubic Myers term vanishes on the classical moduli action since the $\Phi^I$ are diagonal matrices. The bosonic mass terms then generate masses of order $\mu$ for the various moduli fields $\vec{\phi}_a$ (again breaking the $SO(9)$ global symmetry), which we estimated as,
\begin{eqnarray}
S^{E,classical}_{BMN}& =& \frac{N}{\lambda} \int d\tau \sum_{a}\left(  \frac{1}{2}  \partial_\mu \vec{\phi}_a \cdot \partial_\mu \vec{\phi}_a + \frac{\mu^2}{2} \sum_{I=1,2,3} (\phi^I_a)^2 + \frac{\mu^2}{8} \sum_{I=4,\ldots,9} (\phi^I_a)^2 + \frac{1}{4} F_{\mu\nu a} F^{\mu\nu}_a  \right) \nonumber \\
&\sim&   \int dx^p \, \frac{N^2}{\lambda \beta} \phi^2 + \frac{N^2 \beta \mu^2}{\lambda} \phi^2
\end{eqnarray} 
ignoring the gauge fields.
Now the requirement that $x \sim O(1)$, i.e. $\beta \mu \sim 1$, precisely means these two terms are of the same parametric order.
Hence the estimate for the classical term is unchanged after the addition of the BMN mass.

We see that strictly the `moduli' fields $\vec{\phi}_a$ are no longer massless moduli, but we shall still refer to them as such as we shall shortly see they are parametrically light compared to the massive off diagonal modes.
Now consider the 1-loop terms. These arise from integrating over the heavy off diagonal modes of the matrices. There are now two mass scales, that going as $| \vec{\phi}_a - \vec{\phi}_b |$ and given by the moduli, and that arising from the explicit mass $\mu$. The heavier of the two mass scales will control the leading behaviour of the 1-loop term. In our estimates above, without mass, we found that for $t \ll 1$ then $\beta | \vec{\phi}_a - \vec{\phi}_b | \gg 1$. Now since $x \sim O(1)$ we have $\mu \beta \sim 1$, and hence we see that $| \vec{\phi}_a - \vec{\phi}_b | \gg \mu$. Thus our massless estimate indicates that the mass induced by the moduli separation is parametrically larger than the explicit BMN mass $\mu$ in the low temperature regime of interest $t \ll 1$. Hence adding in this BMN mass should make no difference to the leading estimate for the 1-loop behaviour of the moduli fields.

We therefore conclude that for fixed $x = \mu / T\sim O(1)$ the power law dependence of energy (and free energy) on temperature, and also the dependence of the scalar vevs on temperature will be as in the massless case. Hence we expect,
\begin{eqnarray}
\epsilon &\sim& A(x) \, N^2 t^{\frac{14}{5}} \lambda^{\frac{1}{3}} \; ,  \nonumber \\
 \sqrt{     \sum_{I=1,2,3} \langle\frac{ \tr \Phi^I \Phi^I }{N} \rangle }   &\sim& B_1(x) \,t^{\frac{2}{5}} \lambda^{\frac{1}{3}} \; , \qquad
 \sqrt{  \sum_{I=4\, \ldots, 9}\langle \frac{ \tr \Phi^I \Phi^I }{N} \rangle } \sim B_2(x) \,t^{\frac{2}{5}} \lambda^{\frac{1}{3}}
 \end{eqnarray}
in the limit $T\, , \mu \ll \lambda^{1/3}$. Comparing to our naive expectation above, we see that our estimate equating tree and 1-loop terms has determined $p(x)$ and $q(x)$ as being constant.
Since the supergravity black holes have not yet been found we may call this a \emph{prediction} for supergravity.\footnote{Currently Costa, Penedones, and Santos \cite{Costa} are numerically constructing these solutions, and it should soon be possible to compare these predictions  with their numerical results.}

%
\section{Computing the 1-loop corrections}
\label{sec:calc}
%

We now detail the calculation to determine the 1-loop corrections to the classical moduli action, the relevant results of which we stated in equations \eqref{eq:quantcorr} and \eqref{eq:quantcorrthermal}. This calculation is a straightforward generalisation of the calculations considered previously in the context of Matrix theory \cite{Banks:1996vh,Okawa:1998pz,Ambjorn:1998zt}, the principle difference being that our moduli are fields rather than being only functions of time, and that we will be interested in calculating both thermal and non-thermal corrections.

Rather than perform a full 1-loop calculation, we will use the shortcut of computing  the 1-loop correction to a simple classical solution, and then use symmetries to deduce the general answer. 
This will not yield the full 1-loop correction but will give us the information we require to make the estimates discussed above.
Formally our calculation requires $p \ge 2$, but since we will keep $p$ general and obtain an answer with analytic dependence on $p$ we may deduce the result for any $p$, and in particular the cases $p<3$ we are interested in here.

We consider off-diagonal fluctuations  $\delta \vec{\Phi}, \delta {A}^\mu, \delta \Psi$ about a classical solution of the moduli fields $a^\mu_a, \vec{\phi}_a$, so,
\begin{eqnarray}
A^\mu_{ab} =  a^\mu_a \delta_{ab} + \delta \tilde{A}^\mu_{ab} \; , \quad \vec{\Phi}_{ab} = \vec{\phi}_a \delta_{ab} +  \delta \vec{\Phi}_{ab}  \; , \quad \Psi_{ab} = \delta \Psi_{ab} \; .
\end{eqnarray}
Since we are not interested in computing the effective action contribution involving derivatives of the gauge moduli, we will take these to be constant and only non-trivial in the Euclidean time direction so that,
\begin{eqnarray}
a^0_a =  a_a \; , \quad a^i_a = 0 
\end{eqnarray}
for constants $a_a$ which encode the holonomy of the $N$ $U(1)$ gauge fields about the time circle. Being non-compact there are no holonomy moduli associated to the spatial directions.
Since the  classical scalar moduli equations of motion are the harmonic conditions $\partial^\mu \partial_\mu \vec{\phi}_a = 0$ a linear coordinate dependence gives a solution and allows us to consider gradients in the scalars. 
An important point is that as Euclidean time is periodic, we may only choose linear dependence in the spatial directions and not in Euclidean time. 
We restrict all the moduli $\vec{\phi}_a$ to have dependence on two spatial directions, which we take to be $x^1$ and $x^2$. Then we have,
\begin{eqnarray}
\vec{\phi}_a =  \vec{q}_a + x^1 \vec{v}_a + x^2 \vec{u}_a 
\end{eqnarray}
where $\vec{q}_a$, $\vec{v}_a$ and $\vec{u}_a$ are constants. Technically this implies that $p \ge 2$. As stated above, since we will keep $p$ general, then in the end we will deduce our answer for general $p$ including the cases $p<3$. 
We will now split the world-volume spatial coordinates $x^i = \{ x^1, x^2, \ldots , x^p \}$ into $x^1$ and $x^2$ and the remainder $x^{\bar{i}} = \{ x^3, \ldots , x^p \}$.

Due to the gauge invariance we must introduce Fadeev-Popov ghost fields and a gauge fixing term. Following \cite{Okawa:1998pz} we choose the standard background field gauge choice for the fluctuations,
\begin{eqnarray}
G \equiv \partial^\mu \delta A_\mu - i \left[ \phi^I , \delta \Phi^I \right] = 0
\end{eqnarray}
adding $\frac{N}{\lambda}  \int d\tau dx^p \, \tr \frac{1}{2} G^2$ to the action.
Having removed the gauge zero modes, the action may be expanded in the fluctuations, the leading term being that from the classical moduli action so,
\begin{eqnarray}
S^{E,classical} = \frac{N}{\lambda}  \int d\tau dx^p \, \frac{1}{2}  \sum_a | v_a |^2 \; .
\end{eqnarray}
There is no linear term since we are expanding about a classical solution, and at quadratic order we find for our bosonic fields, including the gauge fixing term,
\begin{eqnarray}
S^{E}_{B,quad} &=&  \frac{1}{2} \frac{N}{\lambda}  \int d\tau dx^p \sum_{a \ne b} \, \left( \delta A^0_{ba} , \delta A^1_{ba} ,  \delta A^2_{ba} , \delta A^{\bar{i}}_{ba}, \delta \Phi^I_{b a} \right) \cdot \mathbf{M} \cdot  \left( \begin{array}{c} \delta A^0_{ab} \\  \delta A^1_{ab} \\  \delta A^2_{ab} \\ \delta A^{\bar{j}}_{ab} \\ \delta \Phi^J_{ab} \end{array} \right) 
\nonumber \\
\mathbf{M} & = & \left( \begin{array}{ccccc} 
\Delta_{ab} & 0 & 0 & 0 & 0 \\
0 & \Delta_{ab} & 0 & 0 & - 2 i (v^J_{a} - v^J_{b}) \\
0 & 0 & \Delta_{ab} & 0 & - 2 i (u^J_{a} - u^J_{b}) \\
0 & 0 & 0 &\Delta_{ab} \delta_{\bar{i}\bar{j}} & 0 \\
0 &  2 i ( v^I_{a} - v^I_{b}) &  2 i ( u^I_{a} - u^I_{b})  & 0 & \Delta_{ab} \delta_{IJ}
\end{array} \right)
\end{eqnarray}
with,
\begin{eqnarray} 
\Delta_{ab} \equiv  - \partial^\mu \partial_\mu + 2 i \left( a_{a} - a_{b} \right) \partial_\tau + \left( a_{a} - a_{b} \right)^2 + | \vec{\phi}_{a} - \vec{\phi}_{b} |^2 \; .
\end{eqnarray}
For the fermions we obtain,
\begin{eqnarray}
S^E_{F, quad} =  \frac{1}{2} \frac{N}{\lambda}  \int d\tau dx^p \sum_{a \ne b} \, \delta\bar{\Psi}_{b a} \left( 
\gamma'^\mu \partial_\mu - i \gamma'^0 \left( {a}_a - {a}_b \right) - i  \vec{\gamma} \cdot \left( \vec{\phi}_a - \vec{\phi}_b \right)
 \right) \delta\Psi_{ab}
\end{eqnarray}
where the Euclidean gamma matrices $\gamma'^\mu$ are related to the Lorentzian ones as $\gamma'^\tau = i \gamma^t$ and $\gamma'^i = \gamma^i$, and finally we have ghost fields,
\begin{eqnarray}
S^E_{ghost, quad} = \frac{N}{\lambda}  \int d\tau d^p x \sum_{a \ne b} \, \bar{c}_{b a}  \Delta_{ab} c_{ab} \; .
\end{eqnarray}
Integrating out the off-diagonal fluctuations at quadratic order gives one-loop determinants which yield an effective action for the moduli. 
Now consider for each colour pair $(ab)$ the following $2 \times 2$ matrix and its eigenvalues $\lambda_{i,ab}$;
\begin{eqnarray}
\label{eq:detmat}
\left(
\begin{array}{cc}
(\vec{v}_a - \vec{v}_b) \cdot (\vec{v}_a - \vec{v}_b) \quad & (\vec{v}_a - \vec{v}_b) \cdot (\vec{u}_a - \vec{u}_b) \\ 
(\vec{v}_a - \vec{v}_b) \cdot (\vec{u}_a - \vec{u}_b) \quad & (\vec{u}_a - \vec{u}_b) \cdot (\vec{u}_a - \vec{u}_b) 
\end{array}
\right) \cdot \mathbf{e}_{i,ab} = \lambda_{i,ab} \, \mathbf{e}_{i,ab} \; , \quad i=1,2 
\end{eqnarray}
Then the bosonic, fermionic and ghost fluctuations result in effective actions which can be written in terms of these eigenvalues as,
\begin{eqnarray}
S^E_{B, quad} &=& - \sum_{a<b} \tr \left[ 6 \ln \left( \triangle_{ab} \right) + \sum_{i=1,2} \ln  \left(  \triangle_{ab} + 2 \sqrt{\lambda_{i,ab}} \right)  + \sum_{i=1,2} \ln \left(  \triangle_{ab} - 2 \sqrt{ \lambda_{i,ab}}  \right)   \right]
\nonumber \\
S^E_{F, quad} &=&  + \sum_{a<b} \tr \Bigg[ 2 \ln  \left(  \triangle_{ab} + \sqrt{\lambda_{1,ab}} + \sqrt{\lambda_{2,ab}} \right) + 2 \ln  \left(  \triangle_{ab} + \sqrt{\lambda_{1,ab}} - \sqrt{\lambda_{2,ab}} \right)  \nonumber \\
&& \qquad \qquad \qquad
+ 2 \ln  \left(  \triangle_{ab} - \sqrt{\lambda_{1,ab}} + \sqrt{\lambda_{2,ab}} \right) + 2 \ln  \left(  \triangle_{ab} - \sqrt{\lambda_{1,ab}} - \sqrt{\lambda_{2,ab}} \right) 
 \Bigg]
\nonumber \\
S^E_{G, quad} &=& + \sum_{a<b}\tr \left[  2 \ln \triangle_{ab}  \right]
\end{eqnarray}
respectively, and the trace is taken over momenta, with periodic boundary conditions in time for the bosons and ghosts, and anti-periodic for the fermions. As detailed in Appendix A these traces may be evaluated to yield the 1-loop effective action which consists of colour pairwise interactions,
\begin{eqnarray}
S^{E,1-loop} =  \int d\tau d^p x \sum_{a<b} L\left[ a_a - a_b, \vec{q}_a - \vec{q_b} ,  \vec{v}_a - \vec{v_b},  \vec{u}_a - \vec{u_b}  \right]
\end{eqnarray}
where keeping terms involving up to 4 powers in the gradients, $\vec{v}_a$ and $\vec{u}_a$, then,
\begin{eqnarray}
L\left[ a, \vec{q} ,  \vec{v} ,  \vec{u}  \right] & \equiv & \Bigg[
- \frac{\Gamma\left( \frac{7 - p}{2} \right)}{( 4 \pi )^\frac{1+p}{2}} \frac{ \left( 2 \left(  | \vec{v} |^4 +  | \vec{u} |^4  + 2 \left( \vec{v}.\vec{u} \right)^2 \right) - \left( | \vec{v}|^2 + | \vec{u}|^2  \right)^2 \right)  }{ | \vec{\phi} |^{7-p} }   \nonumber \\
&&  + \frac{e^{-\beta | \vec{\phi} | } \cos\left( \beta a \right)}{( 2 \pi)^\frac{p}{2} \beta}  \left(  \frac{\beta}{| \vec{\phi} |}  \right)^{-\frac{p}{2}} \Bigg[ 
-32 - \frac{4}{3} \left( | \vec{v} |^2 + | \vec{u} |^2 \right) \left( \frac{\beta}{| \vec{\phi} |} \right)^2 - \frac{4}{3}\left( \left( \vec{v} \cdot \vec{\phi} \right)^2 + \left( \vec{u} \cdot \vec{\phi} \right)^2 \right) \left( \frac{\beta}{| \vec{\phi} |} \right)^3  \nonumber \\
&& \qquad \qquad 
 + \left( - \frac{31}{180}  \left( | \vec{v}|^4 + | \vec{u} |^4 + 2 \left( \vec{v}.\vec{u} \right)^2 \right) + \frac{7}{72} \left( | \vec{v} |^2 + | \vec{u}|^2  \right)^2 \right) \left( \frac{\beta}{| \vec{\phi} |} \right)^4 \nonumber \\
 && \qquad \qquad 
+
 \frac{2}{15} \left( | \vec{v}|^2 \left( \vec{v}\cdot\vec{\phi} \right)^2 + | \vec{u}|^2  \left( \vec{u}\cdot\vec{\phi} \right)^2 + 2 \vec{v}\cdot\vec{u} \left( \vec{v}\cdot\vec{\phi} \right) \left( \vec{u}\cdot\vec{\phi} \right) 
\right) \left( \frac{\beta}{| \vec{\phi} |} \right)^5
  \nonumber \\
    && \qquad \qquad 
-  
 \frac{1}{18} \left( | \vec{v} |^2 + | \vec{u} |^2 \right) \left( \left( \vec{v}\cdot\vec{\phi} \right)^2 + \left( \vec{u}\cdot\vec{\phi} \right)^2 \right)  \left( \frac{\beta}{| \vec{\phi} |} \right)^5
  \nonumber \\
  && \qquad \qquad 
-
\frac{1}{36} \left( \left( \vec{v}\cdot\vec{\phi} \right)^2 + \left( \vec{u}\cdot\vec{\phi} \right)^2 \right)^2  \left( \frac{\beta}{| \vec{\phi} |} \right)^6 
 \Bigg] 
 \nonumber \\
&& \left. + \ldots \Bigg] \right|_{\vec{\phi} =  \vec{q} + x^1 \vec{v} + x^2 \vec{u} }
\end{eqnarray}
where the $\ldots$ represent either terms of higher power than four in velocity, or terms that are suppressed relative to those displayed in the limit
 $\beta | \vec{\phi} | \gg 1$ where $\vec{\phi} =  \vec{q} + x^1 \vec{v}  + x^2 \vec{u}$.
We see that the first term going schematically as $\sim (v^4, v^2 u^2, u^4)/ | \vec{\phi} |^{7-p}$ does not explicitly depend on the temperature $\beta^{-1}$ -- this is a \emph{non-thermal} contribution -- whereas the remaining ones explicitly have temperature dependence -- the \emph{thermal} terms -- and are exponentially suppressed in the limit $\beta |\vec{\phi}_a - \vec{\phi}_b  | \gg 1$, which from the estimates of section \ref{sec:argument} we understand as the low temperature limit.

We have computed the classical and leading one loop corrections up to 4 powers in gradients $\vec{v}_a$, $\vec{u}_a$ in the low temperature limit $\beta | \vec{\phi}_a - \vec{\phi}_b | \gg 1$ for a particular classical solution. We may use this result to obtain terms in the classical moduli action, and the leading 1-loop correction, for a general moduli configuration $a(x^\mu)$, $\vec{\phi}(x^\mu)$ using the fact that, $\vec{v}_a = \partial_1 \vec{\phi}_a$ and $\vec{u}_a = \partial_2 \vec{\phi}_a$ and that spatial rotational $SO(p)$ invariance implies,
\begin{eqnarray}
| \vec{v}|^2 + | \vec{u}|^2 & = & \partial_i \vec{\phi} \cdot \partial^i \vec{\phi} \nonumber \\
\left( \vec{v} \cdot \vec{\phi} \right)^2 + \left( \vec{u} \cdot \vec{\phi} \right)^2 & = & \left( \phi \cdot \partial_i \vec{\phi} \right)  \left( \phi \cdot \partial^i \vec{\phi} \right)  \\
| \vec{v} |^4 +  | \vec{u} |^4  + 2 \left( \vec{v}.\vec{u} \right)^2  & = & \left( \partial_i \vec{\phi} \cdot \partial_j \vec{\phi} \right) \left( \partial^i \vec{\phi} \cdot \partial^j \vec{\phi} \right) \nonumber \\
| \vec{v}|^2 \left( \vec{v}\cdot\vec{\phi} \right)^2 + | \vec{u}|^2  \left( \vec{u}\cdot\vec{\phi} \right)^2 + 2 \vec{v}\cdot\vec{u} \left( \vec{v}\cdot\vec{\phi} \right) \left( \vec{u}\cdot\vec{\phi} \right)  & = &  \left( \phi \cdot \partial_i \vec{\phi} \right)  \left( \phi \cdot \partial_j \vec{\phi} \right) \left( \partial^i \vec{\phi} \cdot \partial^j \vec{\phi} \right) \; .
\nonumber 
\end{eqnarray}
Without taking linear dependence on two coordinates (and hence technically requiring that $p\ge2$), but taking only the simpler case of dependence on one, then one would not be able to distinguish all the tensor structures above.

Notice that the derivatives above are spatial $\partial_i$ rather than spacetime $\partial_\mu$ derivatives.
An important point is that we did not consider linear $\tau$ dependence (due to periodic Euclidean time), and because the thermal boundary conditions explicitly break the time-space symmetry, we cannot simply deduce the structure of terms involving $\tau$ derivatives from those involving spatial ones, ie. the effective action is not Lorentz invariant.
An important exception to this is the non-thermal term, which does not explicitly depend on the thermal circle period, and arises also for non-compact Euclidean time. In this case the term is not sensitive to the thermal boundary conditions breaking time-space symmetry and should exhibit the worldvolume Euclidean Lorentz symmetry, and we use this fact to deduce the expression for $L_{non-thermal}$ below in equation \eqref{eq:ansnonthermal}. 
We also emphasise that the thermal potential term which involves no derivatives is fully determined by our analysis. However, generally for the thermal terms it is only terms involving spatial derivatives that we can compute, and our analysis misses those with time derivatives. 

A second important point is that we only computed terms involving single spatial derivatives acting on the scalars. Since we computed these terms by considering fluctuations about an exact solution with the scalars having linear dependence on the spatial coordinates then terms such as $\partial_i \partial_j \vec{\phi}_a$ vanish and hence are not visible to our analysis (obviously unless they are related to our terms by integration by parts). We note that at four derivative order the non-thermal term is highly constrained by supersymmetry \cite{Paban:1998ea,Paban:1998qy,Lowe:1998ch,Sethi:1999qv} and we do not expect such terms involving second derivatives. However, we certainly do expect that for the thermal contribution there will be terms of such a form that we have missed.

Hence we may deduce a covariant form for the 1-loop correction to the moduli action which fully determines the non-thermal terms but only gives certain contributions to the thermal corrections, and ignores derivatives in the gauge fields.
We find,
\begin{eqnarray}
S^{E,1-loop} = \int d\tau d^p x \sum_{a<b} \left( L_{non-thermal} \left[ \vec{\phi}_a - \vec{\phi}_b \right] + e^{-\beta | \vec{\phi}_a -  \vec{\phi}_b  | } \left( \frac{ U_a U^\star_b + U_b U^\star_a }{2}\right) L_{thermal}\left[ \beta , \vec{\phi}_a - \vec{\phi}_b \right] + \ldots \right)  \nonumber \\
\end{eqnarray}
where the $\ldots$ above indicates we are keeping only the leading terms in the limit $\beta | \vec{\phi}_a - \vec{\phi}_b | \gg 1$, and the $U_a$ are holonomies about the time circle,
\begin{eqnarray}
U_a = e^{i \oint d\tau a_a} = e^{i \oint dx^\mu a_{\mu a}} 
\end{eqnarray}
and,
\begin{eqnarray}
\label{eq:ansnonthermal}
L_{non-thermal}\left[ \vec{\phi}  \right] & = & 
- \frac{\Gamma\left( \frac{7 - p}{2} \right)}{( 4 \pi )^\frac{1+p}{2}} \left( 2 \frac{ \left( \partial_\mu \vec{\phi} \cdot \partial_\nu \vec{\phi} \right) \left( \partial^\mu \vec{\phi} \cdot \partial^\nu \vec{\phi} \right) }{ | \vec{\phi} |^{7-p} } -  \frac{\left( \partial_\mu \vec{\phi} \cdot \partial^\mu \vec{\phi} \right)^2 }{ | \vec{\phi} |^{7-p} }  \right) + \mathrm{higher\;derivatives}
\nonumber \\
L_{thermal}\left[ \beta ,  \vec{\phi}  \right] & \supset & 
 \frac{1}{( 2 \pi)^\frac{p}{2} \beta}  \left(  \frac{\beta}{| \vec{\phi} |}  \right)^\frac{p}{2} \left( 
-32 - \frac{4}{3}  \left( {\partial}_i \vec{\phi} \cdot {\partial}^i \vec{\phi} \right) \left( \frac{\beta}{| \vec{\phi} |} \right)^2 
- \frac{4}{3}  \left( \vec{\phi} \cdot \partial_i \vec{\phi} \right) \left( \vec{\phi} \cdot \partial^i \vec{\phi} \right) \left( \frac{\beta}{| \vec{\phi} |} \right)^3 \right.  \nonumber \\
&&
\qquad  \qquad- \frac{31}{180}   \left( \partial_i \vec{\phi} \cdot \partial_j \vec{\phi} \right) \left( \partial^i \vec{\phi} \cdot \partial^j \vec{\phi} \right)  \left( \frac{\beta}{| \vec{\phi} |} \right)^4
+ \frac{7}{72}  \left( \partial_i \vec{\phi} \cdot \partial^i \vec{\phi} \right)^2 \left( \frac{\beta}{| \vec{\phi} |} \right)^4
\nonumber \\
&&
\qquad  \qquad+ \frac{2}{15}   
\left( \phi \cdot \partial_i \vec{\phi} \right)  \left( \phi \cdot \partial_j \vec{\phi} \right) \left( \partial^i \vec{\phi} \cdot \partial^j \vec{\phi} \right)
  \left( \frac{\beta}{| \vec{\phi} |} \right)^5
 \nonumber \\
 &&
 \qquad  \qquad- \frac{1}{18}  \left( \partial_i \vec{\phi} \cdot \partial^i \vec{\phi} \right) \left( \vec{\phi} \cdot \partial_j \vec{\phi} \right) \left( \vec{\phi} \cdot \partial^j \vec{\phi} \right)  \left( \frac{\beta}{| \vec{\phi} |} \right)^5
 \nonumber \\
 &&
\qquad  \qquad
\left. - \frac{1}{36}  \left( \left( \vec{\phi} \cdot \partial_i \vec{\phi} \right) \left( \vec{\phi} \cdot \partial^i \vec{\phi} \right) \right)^2  \left( \frac{\beta}{| \vec{\phi} |} \right)^6 + \mathrm{higher\;derivatives}
 \right)  
\end{eqnarray}
where again we emphasise the derivatives $\partial_i$ in the thermal terms given involve only the spatial directions, and not time, whereas we expect the non-thermal term to be (Euclidean) Lorentz invariant, hence its spacetime derivatives $\partial_\mu$ above.

It would be interesting to compute the 1-loop thermal corrections about a general moduli configuration, rather than the particular one we have chosen, to confirm that the thermal terms we have missed, those involving second or higher derivatives, and those involving time derivatives, are indeed exponentially suppressed as we assume here. For completeness it would be satisfying to also compute the derivative terms in the gauge fields.

%
\section{Summary}
\label{sec:discussion}
%

This paper has been concerned with $(1+p)$-dimensional MSYM at finite temperature in the 't Hooft limit, where the dimensionless temperature $t = T / \lambda^{1/(3-p)}$ is taken to be finite as $N \to \infty$. For low temperatures $t \ll 1$ Maldacena duality conjectures the thermal physics is described by IIA or IIB supergravity black holes. 
We have discussed the low energy moduli of this theory and their 1-loop quantum corrections, focussing on the scalar moduli.
Having computed these 1-loop corrections we then argue, following the logic applied for $p=0$ in the Matrix theory limit \cite{Horowitz:1997fr,Li:1997iz,Banks:1997tn} and more recently by Smilga in the 't Hooft limit we consider here \cite{Smilga:2008bt}, that a new scale emerges when this loop expansion becomes strongly coupled, which we estimate by equating the classical and 1-loop terms in the moduli effective action. 
Under reasonable assumptions these estimates yield precisely the correct temperature and $N$ power law behaviour for both the energy density and scalar vevs that are predicted by the dual supergravity black holes. Our canonical approach naturally reproduces both the $N$ and $t$ dependence, whereas the $p=0$ Hamiltonian analysis in \cite{Smilga:2008bt} assumed the correct $N$ dependence in deriving the behaviour of $t$. 

The simple power law predictions of supergravity only apply when it is valid, in the low temperature regime where $t \ll 1$.
Our canonical approach allows us to qualify the 1-loop corrections into non-thermal and thermal components. We find that the simple scalings predicted by supergravity arise from equating the  non-thermal corrections (those that arise in the zero temperature limit) of lowest derivative order at 1-loop with the classical moduli action. Self consistency of ignoring the thermal corrections and higher derivative non-thermal terms requires that one is in the low temperature regime, so the dimensionless temperature $t \ll1$. Thus not only are we able to reproduce the supergravity predictions, but we also reproduce their regime of validity from the picture of having strongly coupled 1-loop corrections. For $t \sim 1$ the thermal and higher derivative non-thermal corrections would all contribute equally, and a much more complicated dependence of energy density and scalar vevs on temperature would result, rather than simple power laws.

The arguments we make are based on simple estimates and are therefore imprecise in nature. However, since for all $p < 3$ the supergravity predictions for the behaviour of both energy density and scalars with $N$ and $t$ are reproduced correctly and only in the $t \ll 1$ regime where they should apply, we are compelled to believe this analysis gives the correct account for the origin of the various power law scalings of these physical quantities, and it is not simply numerology. 

%
%
An important conceptual point is that we have considered the low energy moduli of the theory, which number $\sim N$, but when equating the classical and 1-loop terms we estimate a free energy of order $N^2$. Conventionally for a weakly interacting system of $N$ degrees of freedom one might expect a free energy going as $N$, and thus this is somewhat surprising. 
One part of this is the use of the 't Hooft coupling at large $N$, rather than the Yang-Mills coupling.
It also seems an important factor is that the 1-loop   term is a pairwise interaction between all the D-branes.
Understanding more clearly why this $N^2$ dependence arises from our perspective of considering $N$ degrees of freedom is an important open question.

We have provided an application of these ideas to deduce constraints on the thermodynamic behaviour of the IIA supergravity black holes dual to the BMN quantum mechanics in the 't Hooft regime. The BMN theory is a mass deformation of the $p=0$ MSYM. The analysis shows that for fixed ratio of mass to temperature, then the power law scaling of the free energy and scalar vevs with varying temperature in the low temperature limit should be the same as that of the undeformed $p=0$ case (the BFSS theory). Naively one might have imagined that this power law would depend on the value of the fixed dimensionless ratio of mass to temperature, whereas our analysis indicates that it does not. The dual
black holes are metrics of cohomogeneity two, and hence will require numerical methods (such as those reviewed in \cite{Wiseman:2011by}) to find them. We understand such a numerical study will soon be complete and it will be interesting to see whether our prediction is confirmed \cite{Costa}. If it is, then an obvious question is how this constraint on the thermodynamic behaviour can be understood from the supergravity directly.

Obviously one important open question is to confirm  that the scalings we see from equating the classical and 1-loop non-thermal terms are compatible with higher loop orders. Smilga \cite{Smilga:2008bt} has pointed out that for $p=0$ the higher loop corrections   \cite{Becker:1997wh,Becker:1997xw} appear to be compatible with the scaling. We expect that the same will be true for all $p$. It may be that the underlying principle that controls this emergent strong coupling scale is the \emph{generalised conformal symmetry} of Jevicki and Yoneya \cite{Jevicki:1998ub}. Given that for $p=3$ it is the conformal symmetry that dictates the power law scaling of all quantities with temperature, it seems reasonable that this generalised conformal symmetry is really the structure that controls the non-trivial temperature scaling of quantities for $p<3$. Thus understanding the role of this generalised conformal symmetry seems likely to be a fruitful direction to pursue.

As outlined in the introduction, the motivation for this work is to provide an avenue to improve direct computations for MSYM in order to simulate quantum gravity. Our analysis gives insight into the scales in the low energy dynamics relevant for the dual IIA or IIB gravity regimes involving black holes, and we are hopeful that this new information -- in particular the estimates indicating the behaviour of the scalars and their derivatives with temperature -- may allow improvements in the direct solution methods employed. In the analytic context of the mean field theory approach of Kabat, Lifschytz and Lowe \cite{Kabat:1999hp,Kabat:2000zv,Kabat:2001ve,Lin:2013jra} it may inform improved Gaussian approximation schemes. In the Monte Carlo approach \cite{Hanada:2007ti,Catterall:2007fp,Anagnostopoulos:2007fw,Catterall:2008yz} it may allow resources to be focussed more efficiently on the most relevant field configurations.

%
\section*{Acknowledgements}
%

We are very grateful to David Berenstein, Andrew Hickling, Arttu Rajantie, Arkady Tseytlin and Ben Withers for useful comments and discussions.

\appendix

%
\section{Appendix: Evaluating the 1-loop determinants}
\label{sec:details}
%

At 1-loop for each colour pair we must evaluate a sum of terms involving traces over the discrete momentum modes in periodic Euclidean time, and the continuum of spatial momenta, that each take the form,
\begin{eqnarray}
D(M) \equiv \tr\left[  \ln \left( - \partial^\nu \partial_\nu + 2 i a \partial_\tau + a^2 + | \vec{\phi} |^2 + M  \right) \right] 
\end{eqnarray}
for constants $a$, $\vec{q}$, $\vec{v}$, $\vec{u}$  and $M$, and, $\vec{\phi} = \vec{q} + x^1 \vec{v} + x^2 \vec{u}$.
Then for a boson (periodic in Euclidean time) and a fermion (anti-periodic in Euclidean time) we may expand in momenta in time as,
\begin{eqnarray}
D_{boson}(M) &=& \sum_{n \in \mathbb{Z}} {\tr}' \left[  \ln \left( - \partial^{i} \partial_{i} + \left( \frac{2 \pi n}{\beta} - a \right)^2 + | \vec{\phi} |^2 + M \right) \right] \nonumber \\
D_{fermion}(M) &=& \sum_{n \in \frac{1}{2} + \mathbb{Z}} {\tr}' \left[  \ln \left( - \partial^{i} \partial_{i} + \left( \frac{2 \pi n}{\beta} - a \right)^2 + | \vec{\phi} |^2 + M \right) \right]
\end{eqnarray}
where $ {\tr}'$ represents the remaining trace over the spatial momenta. Analogously to the eigensystem defined in \eqref{eq:detmat}
we define,
\begin{eqnarray}
\left(
\begin{array}{cc}
\vec{v} \cdot \vec{v} \; & \vec{v} \cdot \vec{u} \\
 \vec{v} \cdot \vec{u}  \; &  \vec{u} \cdot \vec{u} 
\end{array}
\right) \cdot \mathbf{e}_{i} = \lambda_i \mathbf{e}_{i} \; , \quad i=1,2 
\end{eqnarray}
where the eigenvectors $\mathbf{e}_{i}$ corresponding to the (real) eigenvalues $\lambda_{1,2}$ are chosen to be orthonormal (since the matrix is real symmetric). Using this orthonormal basis we define quantities $\alpha_{1,2}$, $A_{1,2}$ and new coordinates $y^{1,2}$ by,
\begin{eqnarray}
\left(
\begin{array}{c}
\vec{q} \cdot \vec{v} \\
 \vec{q} \cdot \vec{u} 
\end{array}
\right) &=& \alpha_1  \mathbf{e}_{1} +  \alpha_2  \mathbf{e}_{2} 
\, , \quad
  \left(
\begin{array}{c}
x^1 \\
x^2
\end{array}
\right) = y^1  \mathbf{e}_{1} +  y^2  \mathbf{e}_{2}  
\, , \quad
A_{1,2} = \frac{1}{\lambda_{1,2}} \left( \mathbf{e}^T_{1,2} \cdot \left(
\begin{array}{c}
\vec{v} \cdot \vec{\phi} \\
\vec{u} \cdot \vec{\phi}
\end{array}
\right) 
\right)^2
\; . \quad
\end{eqnarray}
With these quantities the trace over continuous spatial momenta with the explicit dependence on $x^1$ and $x^2$ may be written using the proper time representation as (see for example \cite{Okawa:1998pz} where there is dependence on one coordinate),
\begin{eqnarray}
D_{boson}(M) &=&  \sum_{n \in \mathbb{Z}}  \int  d^p x \int_0^\infty \frac{d \sigma}{\sigma} \sqrt{ \frac{ 4 \sqrt{ \lambda_1 \lambda_2 } }{ ( 4 \pi)^p \sigma^{p-2} \sinh{2 \sqrt{\lambda_1} \sigma} \sinh{2 \sqrt{\lambda_2} \sigma}  } }  \\
&& \qquad \qquad \qquad \times \exp\left[ - \sigma \left( M + \left( \frac{2 \pi n}{\beta} - a \right)^2 + \left| \vec{\phi} \right|^2  + \sum_{i=1,2} \left( \frac{ \tanh{ \sqrt{\lambda_i} \sigma} }  { \sqrt{\lambda_i} \sigma } - 1 \right) A_i \right) \right] \nonumber 
\end{eqnarray}
and for $D_{fermion}(M)$ we have the same with $\sum_{n \in \mathbb{Z}} \to \sum_{n \in \frac{1}{2} + \mathbb{Z}}$.

For each colour pair, writing $\Delta \equiv  - \partial^\mu \partial_\mu + 2 i a \partial_\tau + a^2 + | \vec{q} + x^1 \vec{v} + x^2 \vec{u} |^2$  then the boson plus ghost effective action, $D^{B+G}$, takes the form, 
\begin{eqnarray}
D^{B+G} &=&  - \tr \left[  4 \ln \left( \triangle \right) + \sum_{i=1,2} \ln  \left(  \triangle + 2 \sqrt{ \lambda_i } \right)  + \ln \left(  \triangle - 2 \sqrt{ \lambda_i } \right)  \right] 
 \\ 
& = & - 4 D_{boson}(0) - D_{boson}(2  \sqrt{\lambda_1} ) - D_{boson}(-2  \sqrt{\lambda_1} ) - D_{boson}(2  \sqrt{\lambda_2}) - D_{boson}(-2  \sqrt{\lambda_2})
\nonumber \\ 
\nonumber \\
& = & -  \sum_{n \in \mathbb{Z}}  \int  d^p x \int_0^\infty \frac{d \sigma}{\sigma} \sqrt{ \frac{ 4 \sqrt{ \lambda_1 \lambda_2 } }{ ( 4 \pi)^p \sigma^{p-2} \sinh{2 \sqrt{\lambda_1} \sigma} \sinh{2 \sqrt{\lambda_2} \sigma}  } }  \left( 4 + 2 \cosh 2 \sqrt{\lambda_1} \sigma + 2 \cosh 2  \sqrt{\lambda_2} \sigma \right)\nonumber \\
&& \qquad \qquad \qquad \times \exp\left[ - \sigma \left( M + \left( \frac{2 \pi n}{\beta} - a \right)^2 + \left| \vec{\phi} \right|^2  + \sum_{i=1,2} \left( \frac{ \tanh{ \sqrt{\lambda_i} \sigma} }  { \sqrt{\lambda_i} \sigma } - 1 \right) A_i \right) \right]
\; . \nonumber
 \end{eqnarray}
For the fermions we have, 
 \begin{eqnarray}
D^{F} &=& 
2 \tr \left[  \ln  \left(  \triangle + \sqrt{ \lambda_1 } + \sqrt{ \lambda_2 } \right)  + 
\ln  \left(  \triangle + \sqrt{ \lambda_1 } - \sqrt{ \lambda_2 } \right) +
 \ln  \left(  \triangle - \sqrt{ \lambda_1 } + \sqrt{ \lambda_2 } \right) +
 \ln  \left(  \triangle - \sqrt{ \lambda_1 } - \sqrt{ \lambda_2 } \right)
 \right]  \nonumber \\ 
& = & 2 D_{fermion}(\sqrt{\lambda_1}+\sqrt{\lambda_2} ) + 2 D_{fermion}(\sqrt{\lambda_1}-\sqrt{\lambda_2} )  + 2 D_{fermion}(-\sqrt{\lambda_1}+\sqrt{\lambda_2} )  \nonumber \\
&& \qquad \qquad\qquad+ 2 D_{fermion}(-\sqrt{\lambda_1}-\sqrt{\lambda_2} ) 
 \\ 
\nonumber \\
& = & +  \sum_{n \in \frac{1}{2} + \mathbb{Z}}  \int  d^p x \int_0^\infty \frac{d \sigma}{\sigma} \sqrt{ \frac{ 4 \sqrt{ \lambda_1 \lambda_2 } }{ ( 4 \pi)^p \sigma^{p-2} \sinh{2 \sqrt{\lambda_1} \sigma} \sinh{2 \sqrt{\lambda_2} \sigma}  } }  \left( 8 \cosh \sqrt{\lambda_1} \sigma  \cosh  \sqrt{\lambda_2} \sigma \right)\nonumber \\
&& \qquad \qquad \qquad \times \exp\left[ - \sigma \left( M + \left( \frac{2 \pi n}{\beta} - a \right)^2 + \left| \vec{\phi} \right|^2  + \sum_{i=1,2} \left( \frac{ \tanh{ \sqrt{\lambda_i} \sigma} }  { \sqrt{\lambda_i} \sigma } - 1 \right) A_i \right) \right]
\; . \nonumber
 \end{eqnarray}
We may expand these for small gradients in powers of $\vec{v}$ and $\vec{u}$, which translates into expanding the above expression in powers of the eigenvalues $\lambda_{1,2}$ (noting that $A_i$ are $O(1)$ in the limit of small $\vec{v}$ and $\vec{u}$). 
Working to the order appropriate for four powers in the gradients $\vec{v}$ and $\vec{u}$ we obtain,
\begin{eqnarray}
 D^{B+G}
&=&
\sum_{n \in \mathbb{Z}}  \int  d^p x \int_0^\infty \frac{d \sigma}{( 4 \pi)^{p/2} \sigma^{1+\frac{p}{2}}} e^{- \sigma \left(  \left( \frac{2 \pi n}{\beta} - a \right)^2 + \left| \vec{\phi} \right|^2 \right) } \nonumber \\
&& \qquad  \qquad \times \Bigg( -8 - \frac{4}{3} \left( \lambda_1 + \lambda_2 \right) \sigma^2 - \frac{8}{3} \left( \lambda_1 A_1 + \lambda_2 A_2 \right) \sigma^3 + \left( - \frac{4}{5}  \left( \lambda_1 - \lambda_2 \right)^2 + \frac{8}{45} \lambda_1 \lambda_2  \right)\sigma^4 
\nonumber \\
&&  \qquad  \qquad \qquad  + \left(  \frac{16}{15} \left( \lambda_1^2 A_1 + \lambda_2^2 A_2 \right) - \frac{4}{9}  \left( \lambda_1 +\lambda_2 \right) \left( \lambda_1 A_1 + \lambda_2 A_2 \right) \right)\sigma^5   - \frac{4}{9} \left( \lambda_1 A_1 + \lambda_2 A_2 \right)^2 \sigma^6 
\nonumber \\
&&  \qquad  \qquad \qquad \qquad\qquad
+ \ldots \Bigg)
\nonumber \\
D^{F}
&=&
\sum_{n \in \frac{1}{2} + \mathbb{Z}}  \int  d^p x \int_0^\infty \frac{d \sigma}{( 4 \pi)^{p/2} \sigma^{1+\frac{p}{2}}} e^{- \sigma \left(  \left( \frac{2 \pi n}{\beta} - a \right)^2 + \left| \vec{\phi} \right|^2 \right) } \nonumber \\
&& \qquad  \qquad \times \Bigg(  + 8 + \frac{4}{3} \left( \lambda_1 + \lambda_2 \right) \sigma^2 + \frac{8}{3} \left( \lambda_1 A_1 + \lambda_2 A_2 \right) \sigma^3 + \left( - \frac{1}{5}  \left( \lambda_1 - \lambda_2 \right)^2 - \frac{8}{45} \lambda_1 \lambda_2  \right)\sigma^4 
\nonumber \\
&&  \qquad  \qquad \qquad  + \left( - \frac{16}{15} \left( \lambda_1^2 A_1 + \lambda_2^2 A_2 \right) + \frac{4}{9}  \left( \lambda_1 +\lambda_2 \right) \left( \lambda_1 A_1 + \lambda_2 A_2 \right) \right)\sigma^5   + \frac{4}{9} \left( \lambda_1 A_1 + \lambda_2 A_2 \right)^2 \sigma^6 
\nonumber \\
&&  \qquad  \qquad \qquad \qquad\qquad
+ \ldots \Bigg)
\end{eqnarray}
where the $\ldots$ represent terms involving higher powers of $\vec{v}$ and $\vec{u}$. Now we perform the $\sigma$ integration, use the relations,
\begin{eqnarray}
\lambda_1 + \lambda_2 & = & \vec{v}\cdot\vec{v} + \vec{u}\cdot\vec{u} 
\; , \qquad  \lambda_1 A_1 + \lambda_2 A_2 = \left( \vec{v}\cdot\vec{\phi} \right)^2 + \left( \vec{u}\cdot\vec{\phi} \right)^2 
\; , \qquad \nonumber \\
 ( \lambda_1)^2 A_1 +( \lambda_2)^2 A_2 &=& \vec{v}\cdot\vec{v} \left( \vec{v}\cdot\vec{\phi} \right)^2 + \vec{u}\cdot\vec{u}  \left( \vec{u}\cdot\vec{\phi} \right)^2 + 2 \vec{v}\cdot\vec{u} \left( \vec{v}\cdot\vec{\phi} \right) \left( \vec{u}\cdot\vec{\phi} \right)
 \nonumber \\
\left( \lambda_1 - \lambda_2 \right)^2 & = & 2 \left( \left( \vec{v}\cdot\vec{v} \right)^2 + \left( \vec{u}\cdot\vec{u} \right)^2 + 2 \left( \vec{v}.\vec{u} \right)^2 \right) - \left( \vec{v}\cdot\vec{v} + \vec{u}\cdot\vec{u}  \right)^2
 \nonumber \\
2 \lambda_1 \lambda_2 & = & - \left( \left( \vec{v}\cdot\vec{v} \right)^2 + \left( \vec{u}\cdot\vec{u} \right)^2 + 2 \left( \vec{v}.\vec{u} \right)^2 \right) + \left( \vec{v}\cdot\vec{v} + \vec{u}\cdot\vec{u}  \right)^2
\end{eqnarray}
and define, $D^+ \equiv D^{B+G}$,  $D^- \equiv D^{F}$, together with,
\begin{eqnarray}
H^+_q& \equiv  &\sum_{n \in \mathbb{Z}} \frac{\Gamma\left( q - \frac{p}{2} \right)}{ \left( \left( \frac{2 \pi n}{\beta} - a \right)^2 + \left| \vec{\phi} \right|^2 \right)^{q - \frac{p}{2} }} \; , \quad H^-_q \equiv  \sum_{n \in \frac{1}{2} + \mathbb{Z}} \frac{\Gamma\left( q - \frac{p}{2}  \right)}{ \left( \left( \frac{2 \pi n}{\beta} - a \right)^2 + \left| \vec{\phi} \right|^2 \right)^{q - \frac{p}{2} }} \; .
\end{eqnarray} 
to yield,
\begin{eqnarray}
 D^{\pm} & = & 
 \frac{1}{( 4 \pi)^{p/2}}  \int  d^p x \Bigg(  \mp 8 H^\pm_0  \mp \frac{4}{3} \left( | \vec{v} |^2 + | \vec{u} |^2 \right) H^\pm_2 \mp \frac{8}{3} \left( \left( \vec{v} \cdot \vec{\phi} \right)^2 + \left( \vec{u} \cdot \vec{\phi} \right)^2 \right) H^\pm_3 \nonumber \\
 && \qquad \qquad 
\left( \mp \frac{3}{10}  - \frac{1}{2} \right) \left( 2 \left( \left( \vec{v}\cdot\vec{v} \right)^2 + \left( \vec{u}\cdot\vec{u} \right)^2 + 2 \left( \vec{v}.\vec{u} \right)^2 \right) - \left( \vec{v}\cdot\vec{v} + \vec{u}\cdot\vec{u}  \right)^2 \right) H^\pm_4 \nonumber \\
 && \qquad \qquad 
 \pm \frac{8}{90} \left( - \left( \left( \vec{v}\cdot\vec{v} \right)^2 + \left( \vec{u}\cdot\vec{u} \right)^2 + 2 \left( \vec{v}.\vec{u} \right)^2 \right) + \left( \vec{v}\cdot\vec{v} + \vec{u}\cdot\vec{u}  \right)^2 \right) H^\pm_4 \nonumber \\
  && \qquad \qquad 
\pm  
 \frac{16}{15} \left( \vec{v}\cdot\vec{v} \left( \vec{v}\cdot\vec{\phi} \right)^2 + \vec{u}\cdot\vec{u}  \left( \vec{u}\cdot\vec{\phi} \right)^2 + 2 \vec{v}\cdot\vec{u} \left( \vec{v}\cdot\vec{\phi} \right) \left( \vec{u}\cdot\vec{\phi} \right)
\right) H^\pm_5
  \nonumber \\
    && \qquad \qquad 
\mp  
 \frac{4}{9} \left( \vec{v}\cdot\vec{v} + \vec{u}\cdot\vec{u} \right) \left( \left( \vec{v}\cdot\vec{\phi} \right)^2 + \left( \vec{u}\cdot\vec{\phi} \right)^2 \right) H^\pm_5
  \nonumber \\
  && \qquad \qquad 
\mp
\frac{4}{9} \left( \left( \vec{v}\cdot\vec{\phi} \right)^2 + \left( \vec{u}\cdot\vec{\phi} \right)^2 \right)^2 H^\pm_6
 + \ldots \Bigg) 
\end{eqnarray}
As detailed in the Appendix B the infinite sums $H^{\pm}_q$ can be written as,
\begin{eqnarray}
H^{\pm}_q &=&  \frac{\Gamma\left( q -\frac{p}{2} - \frac{1}{2} \right)}{ \sqrt{4 \pi }} \beta | \vec{\phi} | ^{1 - 2 q + p}  \\
&& \qquad \qquad \qquad 
+ \frac{2^{\frac{3}{2} - q + \frac{p}{2} } \beta^{2 q - p} }{\sqrt{\pi}} \sum_{n = 1}^{\infty} \left( \pm 1 \right)^{n} \left( \frac{n}{\beta | \vec{\phi} |} \right)^{q - \frac{p}{2} - \frac{1}{2}} K_{\frac{1}{2} - q - \frac{p}{2}}\left( n \beta | \vec{\phi} | \right) \cos\left( n \beta a \right)  \; . \nonumber 
\end{eqnarray}
so that in the limit $\beta | \vec{\phi} | \gg 1$ we may expand the modified Bessel function to obtain,
\begin{eqnarray}
 H^{\pm}_q & = &  \frac{\Gamma\left( q - \frac{p}{2} - \frac{1}{2} \right)}{ \sqrt{4 \pi}} \beta | \vec{\phi} | ^{1 - 2 q + p} + O\left( e^{- \beta | \vec{\phi} | } \right) \nonumber \\
 H^{+}_q - H^{-}_q  & = &  2^{2 - q + \frac{p}{2}} e^{- \beta | \vec{\phi} | } \cos\left( \beta a \right) \left( \frac{\beta}{| \vec{\phi} |} \right)^{q - \frac{p}{2}} \left( 1 + O \left( \frac{1}{ \beta | \vec{\phi} | } \right) \right)
\end{eqnarray}
and so using $\int d\tau = \beta$ we arrive at,
\begin{eqnarray}
 D^{B+G} &+& D^{F}  =  D^+ + D^- \nonumber \\
& = &
 \int d\tau \int d^p x \Bigg[ 
- \frac{\Gamma\left( \frac{7 - p}{2} \right)}{( 4 \pi )^\frac{1+p}{2}} \frac{ \left( 2 \left( \left( \vec{v}\cdot\vec{v} \right)^2 + \left( \vec{u}\cdot\vec{u} \right)^2 + 2 \left( \vec{v}.\vec{u} \right)^2 \right) - \left( \vec{v}\cdot\vec{v} + \vec{u}\cdot\vec{u}  \right)^2 \right) }{ | \vec{\phi} |^{7-p} }   \nonumber \\
&&  + \frac{e^{-\beta | \vec{\phi} | }}{( 2 \pi)^\frac{p}{2} \beta}  \cos\left( \beta a \right) \left(  \frac{\beta}{| \vec{\phi} |}  \right)^{-\frac{p}{2}} \Bigg[
-32  - \frac{4}{3} \left( | \vec{v} |^2 + | \vec{u} |^2 \right) \left( \frac{\beta}{| \vec{\phi} |} \right)^2 - \frac{4}{3} \left( \left( \vec{v} \cdot \vec{\phi} \right)^2 + \left( \vec{u} \cdot \vec{\phi} \right)^2 \right) \left( \frac{\beta}{| \vec{\phi} |} \right)^3 \nonumber \\
 && \qquad \qquad 
  + \left( - \frac{31}{180}  \left( \left( \vec{v}\cdot\vec{v} \right)^2 + \left( \vec{u}\cdot\vec{u} \right)^2 + 2 \left( \vec{v}.\vec{u} \right)^2 \right) + \frac{7}{72} \left( \vec{v}\cdot\vec{v} + \vec{u}\cdot\vec{u}  \right)^2 \right) \left( \frac{\beta}{| \vec{\phi} |} \right)^4 \nonumber \\
 && \qquad \qquad 
+
 \frac{2}{15} \left( \vec{v}\cdot\vec{v} \left( \vec{v}\cdot\vec{\phi} \right)^2 + \vec{u}\cdot\vec{u}  \left( \vec{u}\cdot\vec{\phi} \right)^2 + 2 \vec{v}\cdot\vec{u} \left( \vec{v}\cdot\vec{\phi} \right) \left( \vec{u}\cdot\vec{\phi} \right) 
\right) \left( \frac{\beta}{| \vec{\phi} |} \right)^5
  \nonumber \\
    && \qquad \qquad 
-  
 \frac{1}{18} \left( \vec{v}\cdot\vec{v} + \vec{u}\cdot\vec{u} \right) \left( \left( \vec{v}\cdot\vec{\phi} \right)^2 + \left( \vec{u}\cdot\vec{\phi} \right)^2 \right)  \left( \frac{\beta}{| \vec{\phi} |} \right)^5
  \nonumber \\
  && \qquad \qquad 
-
\frac{1}{36} \left( \left( \vec{v}\cdot\vec{\phi} \right)^2 + \left( \vec{u}\cdot\vec{\phi} \right)^2 \right)^2  \left( \frac{\beta}{| \vec{\phi} |} \right)^6 
 \Bigg] 
 \nonumber \\
 && \qquad  + \ldots
\Bigg]
\nonumber
\end{eqnarray}
where we have worked to fourth power in the gradients $\vec{v}$, $\vec{u}$ and at each order in these gradients the terms given by the dots $\ldots$ are subleading in the limit that $\beta | \vec{\phi} | \gg 1$.

%
\section{Appendix: Evaluating an infinite sum}
\label{sec:sum}
%

We now consider the infinite sum,
\begin{eqnarray}
F(b) = \sum_{n \in \mathbb{Z}} \frac{1}{ \left( \left( n - b \right)^2 + M^2 \right)^{w}} \; .
\end{eqnarray} 
Since this is periodic in $b \sim b + 1$ then we may express  it as a Fourier sum,
\begin{eqnarray}
F(b) = \sum_{k = -\infty}^{+\infty} e^{2 \pi i k b} F_k 
\end{eqnarray} 
with Fourier coefficients,
\begin{eqnarray}
F_k & =& \int_0^1 db \, e^{-2 \pi i k b} F(b) = \int_0^1 db  \sum_{n \in \mathbb{Z}} \frac{e^{-2 \pi i k b} }{ \left( \left( n - b \right)^2 + M^2 \right)^{w}} =   \sum_{n \in \mathbb{Z}}  \int_{n}^{n+1} db\, \frac{e^{-2 \pi i k b} }{ \left( b^2 + M^2 \right)^{w}} \nonumber \\
& = &   \int_{-\infty}^{+\infty} db\, \frac{e^{-2 \pi i k b} }{ \left( b^2 + M^2 \right)^{w}}
\end{eqnarray} 
which evaluate to give,
\begin{eqnarray}
F_0 & = & \sqrt{\pi} M^{1- 2 w} \frac{ \Gamma\left( w - \frac{1}{2} \right) }{ \Gamma\left( w \right) }
\end{eqnarray}
and 
\begin{eqnarray}
F_k = F_{-k} = \frac{2 \pi^w}{\Gamma\left(w\right)} \left( \frac{| k |}{ M} \right)^{w-\frac{1}{2}} K_{\frac{1}{2} - w}\left( 2 \pi M | k | \right) \; , \quad k \ne 0 \; .
\end{eqnarray}
Hence,
\begin{eqnarray}
F(b) = \sqrt{\pi} M^{1- 2 w} \frac{ \Gamma\left( w - \frac{1}{2} \right) }{ \Gamma\left( w \right) } + 2 \sum_{k = 1}^{\infty} \frac{2 \pi^w}{\Gamma\left(w\right)} \left( \frac{| k |}{ M} \right)^{w-\frac{1}{2}} K_{\frac{1}{2} - w}\left( 2 \pi M | k | \right) \cos\left(2 \pi k b\right) 
\end{eqnarray}

%
\bibliographystyle{JHEP}
\bibliography{paperV2}
%

\end{document}